# Decoding the Pulse of Community during Disasters: Resilience Analysis Based on Fluctuations in Latent Lifestyle Signatures within Human Visitation Networks

Junwei Ma[1*], Ali Mostafavi[2]

[1] (Corresponding author) Ph.D. Student, Urban Resilience.AI Lab, Zachry Department of Civil and Environmental Engineering, Texas A&M University, College Station, United States; E-mail: jwma@tamu.edu

[2] Associate Professor, Urban Resilience.AI Lab Zachry Department of Civil and Environmental Engineering, Texas A&M University, College Station, United States; E-mail: amostafavi@civil.tamu.edu

## Abstract

Examining the impact of disasters on life activities of populations is critical for understanding community resilience dynamics, yet it remains insufficiently studied in the existing literature. In this study, we leveraged data from more than 1.2 million anonymized human mobility communications across 30 parishes in Louisiana to construct a temporal network that tracks visitation to places from which we characterized human lifestyle signatures before, during, and after Hurricane Ida in 2021. Utilizing the motif model, we distilled complex human lifestyles into identifiable patterns and clustered them into classes: commute, healthcare, dining out, and youth-oriented lifestyle. We defined two metrics to evaluate disruption and recovery fluctuations in lifestyle patterns during the perturbation period compared to the steady period: 1) frequency (daily number of motifs), and 2) proximity (daily average distance of motifs). The results indicate significant dynamics in lifestyle patterns due to the hurricane, with essential facilities (e.g., healthcare) demonstrating a swift recovery. The study underscores the heterogeneity of locations visited and the necessity of integrating both essential and non-essential facilities into disaster response initiatives. Furthermore, our study reveals sustained changes in lifestyle patterns, highlighting the long-term impact of the hurricane on daily life. These insights demonstrate the significance of examining lifestyle signatures and their fluctuations in evaluating disaster resilience patterns for affected communities. The outcomes of this study are poised to aid emergency managers and public officials to more effectively evaluate and monitor disaster impacts and recovery based on changes in lifestyle patterns in the community.

# 1. Introduction

Natural hazards, such as hurricanes, not only disrupt the physical infrastructure but also have profound effects on the movement patterns and daily lifestyles of populations (Coleman, Liu et al. 2023, Ronco, Tárraga et al. 2023). Human lifestyles can be characterized based on patterns of visitation activities (Ma, Li et al. , Di Clemente, Luengo-Oroz et al. 2018, Fan, Wu et al. 2024), such as visiting grocery stores for food necessities or visiting shopping centers for clothing (Fig.1a). The disruption of lifestyles during hurricanes negatively impacts populations, as they are unable to access different facilities due to disaster-induced disruptions (Fig.1b). The restoration of lifestyles can signal a recovery to normalcy, as the population is better able to comfortably resume their previous movement patterns (Jiang, Yuan et al. 2023). The disruption of lifestyles and the return to pre-disruption lifestyle standards could represent an important facet of the community resilience dynamics during and in the aftermath of disasters. Characterizing these changes is crucial for effective disaster management and the promotion of community resilience.

While the physical impacts of hazards are often visible and quantifiable (Dong, Esmalian et al. 2020, Esmalian, Dong et al. 2021), subtler changes in human behavior and lifestyle patterns remain less understood and thus less addressed in previous community resilience studies. The existing approaches to examining disaster impacts and recovery tend to focus on physical infrastructure and the built environment (Bonczak and Kontokosta 2019, Castaño-Rosa, Pelsmakers et al. 2022), often overlooking the perturbations in life activities. This gap is particularly poignant considering the importance of lifestyle patterns and effects of perturbed lifestyle patterns on population wellbeing (Coleman, Liu et al. 2023). The need for a dynamical and nuanced analysis of lifestyle patterns during disasters is evident, yet methodological approaches to quantifying lifestyle resilience in the context of natural hazards are still in their infancy.

Recognizing this gap, this study employs a network motif analysis of anonymized human mobility data to characterize the temporal dynamics of population lifestyle patterns during Hurricane Ida in 2021. This analysis provides a detailed examination of the shifts in visitations to points of interest (POIs) within 30 parishes in Louisiana, a state situated on the Gulf Coast of the Atlantic Ocean, outlining the disruption and recovery extent of various lifestyle motifs ranging from essential patterns, such as healthcare and commuting, to non-essential activities, like dining out and youth-oriented engagements. In particular, the analysis focuses on answering the following research questions: (RQ1): what are the distinctive patterns in typical lifestyles of populations during normal situations? (RQ2): to what extent do disasters impact lifestyles and how quickly do lifestyle patterns recover? (RQ3): to what extent do the impact and recovery vary across different lifestyle patterns? Through answering these questions, the study offers novel approaches and insights related to the disruption and recovery of human lifestyles in hazard events and emphasizes the importance of examining population activities in evaluating community resilience dynamics. Fig.1 provides a conceptual overview of detecting latent lifestyle signatures in this study.

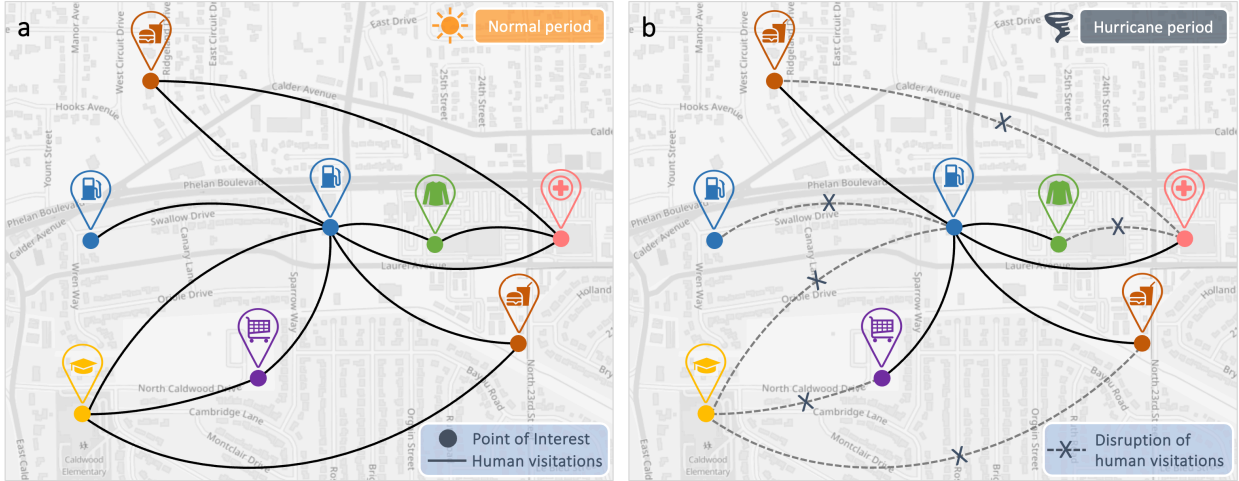

**Fig.1.** Detecting latent lifestyle signatures in human visitation network. **a.** Using human visitation trajectories and POI information, we identified human lifestyle patterns of visitation activities to different categories of locations. **b.** Human visitations are heterogeneously disrupted by the hurricane.

## 2. Background

Lifestyles are shaped by the sequence of daily activities carried out by individuals, often involving visitations to various facilities (e.g., grocery stores, restaurants, and gasoline stations) (Batty 2013). These activities offer valuable insights for urban planning, aiding in decisions related to facility distribution, equity, sustainability, and resilience (Toole, Herrera-Yaqüe et al. 2015, Maeda, Shiode et al. 2019). In recent years, there has been an increasing focus on studying human lifestyles. (Fan, Wu et al. 2024) examined millions of mobile phone communications to deduce important sequences of facilities visited by individuals; they categorized lifestyles according to the patterns observed within each sequence. (Su, McBride et al. 2020) recognized daily activity patterns of students, workers, and commuters during the course of California workdays by using sequences of visits to places in human mobility networks among different regions. (Lei, Chen et al. 2020) analyzed travel patterns of commuting and not-commuting in Nanjing, China, among public transit users. Another stream of studies examined the relationship between urban features and lifestyle patterns. Urban areas are characterized by a variety of facilities, like restaurants, hospitals, and grocery stores, which play a significant role in defining urban functionality and lifestyle patterns (Fusco 2016, Abbiasov, Heine et al. 2024). Despite the growing recognition of the importance of lifestyle patterns in examining urban dynamics, little of the existing work has examined disrupted lifestyle patterns in hazard events for characterizing community resilience dynamics.

Natural hazards (e.g., hurricanes and typhoons) significantly disrupt people's daily lifestyle patterns (Coleman, Liu et al. 2023). While many studies have investigated how lifestyles change in response to hazards, most rely on static and global indicators, such as duration of stay at home (Li and Mostafavi 2022), types of land use visited (Noszczyk, Gorzelany et al. 2022), and mode of transportation used (Thombre and Agarwal 2021). These indicators offer important yet limited perspectives on the complex and evolving dynamics of lifestyles during hazards. Lifestyle patterns are subject to continuous shift with the onset and aftermath of hazards, and static and global measures fail to fully characterize the impacts and recovery of lifestyle patterns. In fact, the interaction between locations and human behavior gives rise to diverse spatial structures that delineate different lifestyle patterns (Ma, Li et al. , Di Clemente, Luengo-Oroz et al. 2018, Fan, Wu et al. 2024). Based on examining the fluctuations in lifestyles during hazards compared with the steady state (normal period), we can evaluate impact and duration of recovery in lifestyle patterns of populations.

To delineate lifestyle patterns and to examine the impact of hazard events on lifestyle patterns and their recovery, we utilized location intelligence data. In recent years, a growing number of studies have utilized location intelligence data for examining community resilience issues like urban flooding (Wang, Loo et al. 2020, Hong, Bonczak et al. 2021), property damage (Kontokosta and Malik 2018, Yuan and Liu 2020), and evacuation strategies (Han, Tsou et al. 2019, Chen, Gong et al. 2020). For example, in the study of Hurricane Harvey's evacuation patterns, (Hong, Bonczak et al. 2021) categorized households based on their response strategies, including sheltering in place, remaining stable, becoming distressed, or abandoning their homes altogether. (Podesta, Coleman et al. 2021) studied the impact of Hurricane Harvey on POIs in Harris County, Texas, highlighting how different POI systems recover at varying rates to their baseline levels. (Esmalian, Coleman et al. 2022, Fan, Jiang et al. 2022) used location intelligence data to assess the equitable accessibility of essential services during disruptive events, identifying distinct patterns in how different user groups access critical facilities. Despite these advancements, there has been limited use of location intelligence data in specifically characterizing the dynamics of lifestyles in the face of natural hazards. Unlike traditional data collection methods, such as surveys (Baker 2011, Josephson, Schrank et al. 2017), location intelligence offers more detailed, larger-scale, and timely data with less burden on the affected population (Ma, Li et al.). This data exhaustively records the movement trajectories of individuals in a way that guarantees privacy (Fan, Wu et al. 2024), not only helping facilitate understanding of the spatial mobility networks but also enabling a more nuanced examination of lifestyles in which people interact with a variety of real-world locations. Despite the growing attention to the value of location intelligence data for characterizing community resilience dynamics, limited attention has been paid to lifestyle impacts and recovery as an important aspect of community resilience.

In this study, we utilized network motifs to characterize and quantity lifestyle signatures. Network subgraphs (motifs) have emerged as an important aspect of examining the dynamics of temporal networks in recent years (Su, McBride et al. 2020, Cao, Li et al. 2021, Yin and Chi 2021, Hsu, Liu et al. 2024). Recent studies show that global network properties (e.g., density and average degree) fail to reveal microscopic perturbations in temporal networks; they propose examining the changes and stability of motifs to better evaluate the characteristics of perturbed temporal networks (Rajput and Mostafavi 2023, Hsu, Liu et al. 2024). Motif is defined as the interconnection mode that has recurrence frequencies in the real network much higher than those in a randomized network (Dey, Gel et al. 2019). (Milo, Shen-Orr et al. 2002) found that motifs are ubiquitous in universal classes of networks, such as biochemical, neurobiological, ecological, and engineering network. As basic building blocks in networks, motifs are crucial for understanding the basic structures that control and modulate many complex system behaviors (Benson, Gleich et al. 2016). Recently, researchers in urban studies have constructed motifs arising from human mobility data, such as bike-sharing ride records (Yang, Heppenstall et al. 2019) and public transportation card swipe records (Lei, Chen et al. 2020), to explore various human movement characteristics. In this study, we examine motifs to delineate lifestyle signatures of communities and their fluctuations to evaluate disaster impacts and recovery. By quantitatively accessing the dynamics in the frequency and proximity metrics of motifs and lifestyle clusters before, during, and after the hazard events, we aim to provide a granular view of lifestyle disruptions and recoveries. The study not only contributes to the existing body of knowledge on urban human activities and community resilience dynamics but also provides insights and methods that can help emergency managers and public officials to effectively examine the impacts and recovery of lifestyles in affected populations to inform resource allocation and prioritization.

## 3. Datasets and methodology

### 3.1 Study area and context

Hurricane Ida, a powerful Category 4 hurricane, originated in the Caribbean Sea in August 23, 2021 and made landfall in Louisiana on August 26, 2021. Hurricane Ida affected multiple coastal regions, notably the New Orleans and Lafayette metropolitan areas with devastating consequences. Hurricane Ida caused numerous direct fatalities and inflicted estimated damage to be more than $100 billion (Caffey, Wang et al. 2022). Our research period spans from August 1, 2021 to September 30, 2021, with the timeframe from August 26 through September 2 designated as the Ida period due to the hurricane's direct impact. This interval is depicted in Fig.3b.

We selected 30 parishes (equivalent to counties) in the New Orleans and Lafayette areas within the primary impact zone of Hurricane Ida's landfall to serve as a crucial testing ground for assessing lifestyle pattern changes. The graphic outline of Ida's path and the 30 specifically selected parishes is provided in Supplementary Fig.1.

### 3.2 Data sources

We analyzed fine-grained human mobility data from Spectus, Inc., a location intelligence and measurement company that collects anonymized and privacy-enhanced mobile phone data (Spectus 2023). The dataset consists of more than 1.2 million anonymized visitations to POIs in the 30 parishes of Louisiana and includes attributes of device ID, POI ID, latitude, longitude, and dwell time of visitations.

The location information of POIs was obtained from SafeGraph, Inc., a location intelligence data company that builds and maintains accurate POI locations for the U.S. (SafeGraph 2023). The dataset includes basic information such as POI ID, location name, address, category, and brand association for POIs where people spend time and money. We labeled each POI with NAICS (North American Industry Classification System) category code, which is the standard used by federal statistical agencies in classifying business establishments (Committee). In this study, we used the four-digit NAICS code which represents the industry groups as the categories of POIs.

### 3.3 Methods

Fig.2 shows the overview of the research workflow. We first matched millions of human visitations in the mobility dataset to POIs in the location dataset, allocating a NACIS code to each POI. Then we aggregated these human visitations to create daily networks of places during a span of 61 days before, during, and following the landfall of Hurricane Ida. Following this, we identified nine basic motifs as human lifestyles through the analysis of networks of places. We then categorized the attributed motifs into four main lifestyle clusters. Finally, we developed two metrics to quantitatively evaluate the temporal fluctuations in human lifestyle patterns.

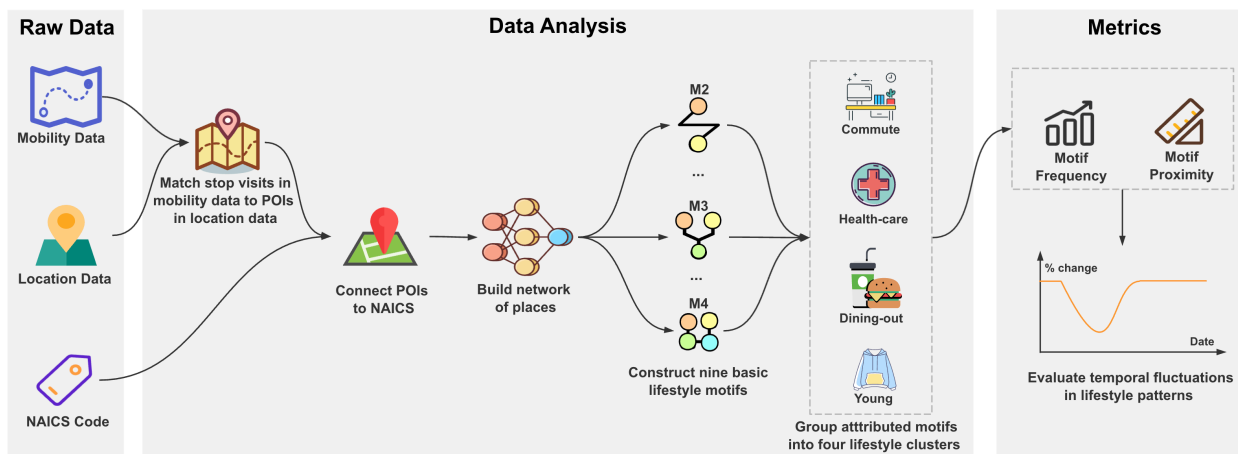

**Fig.2.** Overall workflow of constructing human visitation networks and using network motifs for characterizing lifestyle patterns.

### 3.3.1 Generating visitation network of places

First, we applied the Spectus dataset to identify visits between POIs. The "stop" table from Spectus's core dataset was utilized to pinpoint the POIs visited by a device (i.e., mobile phone user). We employed dwell time of visitation— the length of time devices remain at a POI— as a measure to define a visit. A visit was recorded when a device's stay at a POI exceeded the threshold of 2 minutes. By sequencing the timing of these visits, we determined the origin and destination POIs from which daily aggregations of human visitations between POIs were then compiled. Next, we used the POI IDs in the Spectus dataset to match human visitations between POIs with location data from SafeGraph dataset. Finally, a four-digit NAICS category code was assigned to each POI, adhering to the federal standard for classifying business establishments. To capture a comprehensive daily overview of human visitations, we consolidated the daily visits of all devices into an undirected and weighted network, which reflects the collective patterns and volume of human movement across different POIs in one area. We denoted this network as $G_t$, or the network of places, expressed as

$$G_t = (V, E, W) \qquad (1)$$

where $V$ symbolizes the POIs, $E$ denotes the inter-POI visits, $W$ represents the visit counts between POI pairs, and $t \in [1, 61]$ corresponds to each of the days from August 1 to September 30, 2021. We aggregated the visitations for each day and generated 61 networks of places.

To better distinguish lifestyle patterns, we identified five categories of POIs that provide essential services (health care, grocery stores, gasoline stations, telecommunications carriers, and educational service) and 15 categories of POIs offering non-essential services, based on NAICS code. Supplementary Table 1 summarizes these POI categories. Fig.3c illustrates the proportion of visits to different categories of POIs.

### 3.3.2 Constructing attributed motifs

After creating the visitation networks of places for various days, we extracted the recurring subgraph patterns within these networks, known as motifs. A motif, represented by $G' = (V', E')$, is a recurrent multi-node-induced subgraph pattern within the larger graph $G$, where $V' \subseteq V$ and $E' \in E$, and $E'$ contains all edges $e_{uv} \in E$ such that $u, v \in V'$. Unlike global network properties, such as average degree, density, and diameter, which primarily focus on high-order connectivity features at the level of the whole network, motifs reveal low-order structures and represent the local interaction pattern of the network (Prill, Iglesias et al. 2005). Drawing from prior research (Ma, Li et al. , Hsu, Liu et al. 2024) and computational cost considerations, we selected two-node, three-node, and four-node motifs to portray lifestyle patterns in our analysis, as illustrated in Fig.3a. Motifs such as M2-1, M3-2, M4-1, and M4-2, which are densely interconnected, signify a substructure of POIs reflecting the most integrated lifestyle interactions. In contrast, motifs like M3-1, M4-3, and M4-5 depict more straightforward connections, either in a closed loop or a linear chain. M4-4 includes a triangle of three interconnected nodes and an additional connected node, representing a related but separate location. Meanwhile, M4-6 is illustrative of a hub-and-spoke layout, with a central node linked to three other locations.

However, population lifestyle patterns are not solely captured by the number or configuration of motifs but also by the attributes of the nodes. For example, a two-node motif representing movement from a school to a pharmacy carries different implications than one moving from a grocery store to a shopping mall. Ignoring the attributes of nodes in motifs could lead to a significant loss of information critical for identifying distinct lifestyle patterns. In this study, we associated the nodes within motifs with four-digit NAICS codes to differentiate POI categories and discern diverse lifestyle signatures. Icons representing these POI categories are depicted in Fig.3c, and a breakdown of essential and non-essential POI categories is provided in Supplementary Table 1.

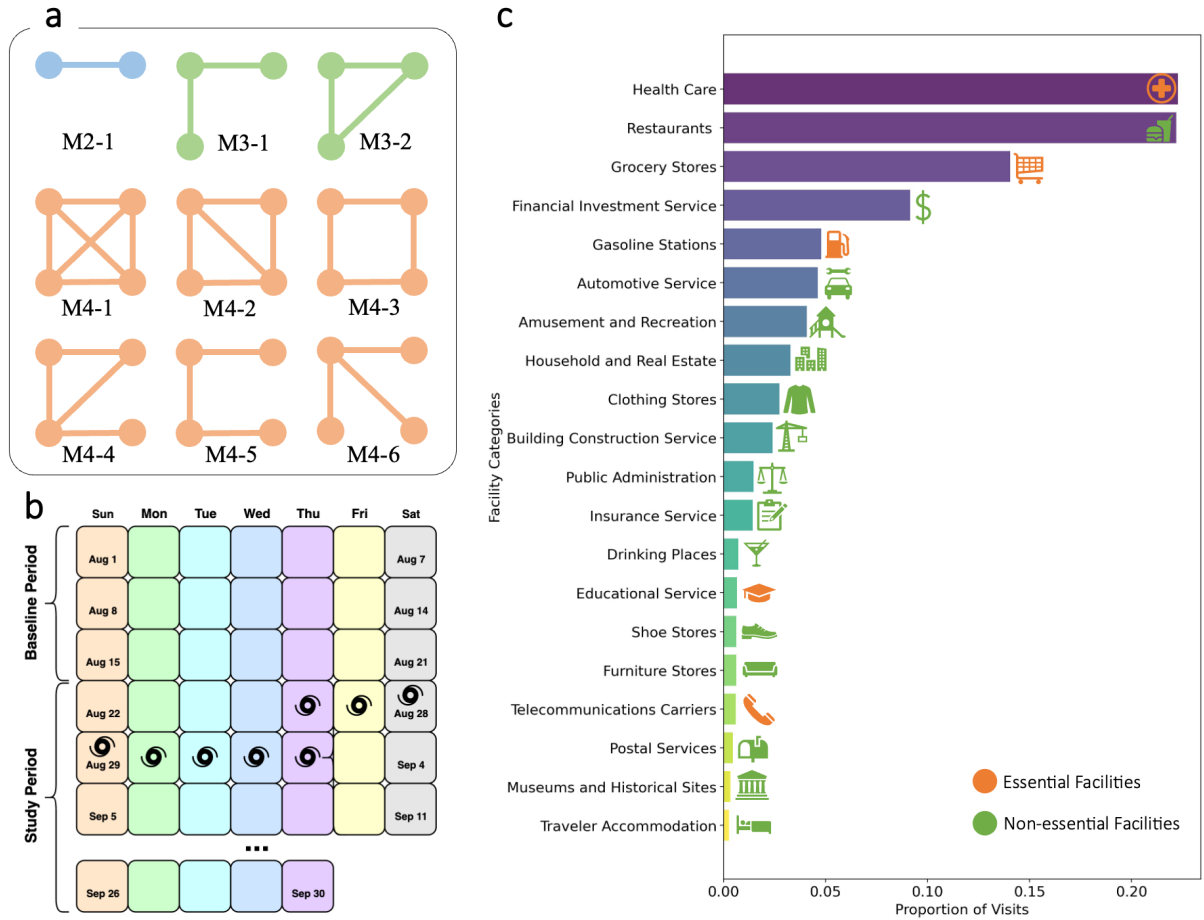

**Fig.3.** Methods for characterizing lifestyle patterns and basic analysis of human visitation. **a.** Motif substructures used in this study. **b.** Baseline setting and study period. **c.** Proportion of visits in different categories of POIs during the study period.

### 3.3.3 Grouping lifestyle clusters

While motifs enriched with node attributes provide a detailed depiction of the population's location-based lifestyles, the sheer diversity of such attributed motifs (despite having grouped POIs into 20 primary categories) makes it impractical to analyze every lifestyle variation during events like hurricanes. Consequently, from the nine basic motifs identified above, we focused on the top ten by quantity within each basic motif for further examination. The fact that the top ten attributed motifs represent over 10% of the totals in their respective categories (Fig.5), justifies concentrating our analysis on them. Further analysis will be detailed in the results section.

Moreover, within these top ten attributed motifs, we noted recurring characterizations of lifestyle patterns. For example, some attributed motifs depict regular visits to offices, indicative of commuting habits, while others center around healthcare facilities, revealing healthcare-related lifestyle tendencies. To further streamline the analysis, we categorized these top ten attributed motifs into four primary lifestyle clusters: commute, healthcare, dining-out, and youth-oriented activities. Fig.6 presents these four principal lifestyle clusters, and Supplementary Table 2 outlines the specific attributed motifs within each cluster.

## 3.4 Metrics

### 3.4.1 Baseline for percentage change calculation

Considering Hurricane Ida's impact on Louisiana beginning on August 26, 2021, we selected the period from August 1 to August 21 as our baseline to avoid the atypical visitations that often precede a hurricane. Previous studies calculated daily average values of metrics such as the number of trips during the baseline period and used these values as a reference for comparisons with the values observed during the study periods (Cox, Prager et al. 2011, Askitas, Tatsiramos et al. 2021). This approach, however, typically fails to account for the weekly lifestyle pattern variations. Ma, Li et al. highlighted significant differences in lifestyle motifs between weekdays and weekends (Ma, Li et al.). To account for weekly fluctuations, we refined the baseline computation in our study as follows: we separated the three-week period by day of the week and averaged the metrics (i.e., the count and the average distance of motifs in the following section) for each corresponding day, Sunday through Saturday. This process resulted in seven baseline values for each metric. Fig.3b provides a diagrammatic explanation of how we set these baselines. The formula for calculating the baselines, denoted as $\bar{p}$, for a particular day is given by:

$$\bar{p}_{ij} = \frac{\sum_{i=1}^{3} p_{ij}}{3}, i = 1, 2, 3; j = 1, 2, \ldots, 7. \tag{2}$$

where, $i$ denotes each of the three weeks, and $j$ corresponds to each day of the week. The variable $p_{ij}$ represents the measured metrics for the $j$-th day of the week during the $i$-th week.

### 3.4.2 Motif frequency

Motif frequency, defined as the count of motifs, serves as an indicator of recurring location-based lifestyle patterns within a temporal frame. To evaluate the impact of the Hurricane Ida on lifestyle patterns, we analyzed changes in motif frequency over time. To facilitate comparison and mitigate scale discrepancies, we calculated the percentage change in motif frequency $c_j$ using the baseline described in the above section as follows:

$$c_j = \frac{c_{nj} - \bar{c}_j}{\bar{c}_j}, n = 22, 23, \ldots, 61; j = 1, 2, \ldots, 7 \tag{3}$$

where $n$ represents the study period from August 22 through September 30, and $j$ represents each day in a week.

### 3.4.3 Motif proximity

Motif proximity refers to the average spatial distance between the nodes within a motif, reflecting the physical closeness of location-based lifestyles. We used motif proximity to measure how lifestyle patterns are spatially distributed. The average distance for motif proximity is computed using the Haversine Formula (Alam, Manaf et al. 2016) to determine the spatial length $d$ of each edge within a motif:

$$d = \frac{\sum_{i,j \in V} 2R \arcsin \sqrt{\sin^2\left(\frac{lat_i - lat_j}{2}\right) + \cos(lat_i)\cos(lat_j)\sin^2\left(\frac{lng_i - lng_j}{2}\right)}}{e} \tag{4}$$

where $R$ is the radius of the Earth, $lat_i, lat_j, lng_i, and\ lng_j$ are the radian coordinates of node $i$ and $j$, and $e$ is the number of edges of a motif.

The percentage change in motif proximity $d_j$ is calculated via the following formula:

$$d_j = \frac{d_{nj} - \bar{d}_j}{\bar{d}_j}, n = 22, 23, \ldots, 61;\ j = 1, 2, \ldots, 7 \tag{5}$$

where *n* represents the study period from August 22 through September 30, and *j* represents each day in a week.

## 4. Results

### 4.1 Lifestyle fluctuations in the visitation network of places

Utilizing Safegraph's location dataset, we labeled each node within the 61 networks of places with the corresponding NAICS code, which enabled us to aggregate and analyze location distribution of lifestyles. Fig.3c ranks the visitation distribution across various POI categories where healthcare facilities and restaurants were the most frequently visited places, each comprising over 20% of all visits. Following these were grocery stores, financial investment services, and gasoline stations, which also saw significant visitation.

We then examined the temporal fluctuations in human movements (i.e., the number of devices and number of flows) in the raw Spectus human mobility data throughout the two-month period of Hurricane Ida, also applying the baseline calculation presented earlier to determine percentage changes. As shown in Fig.4a, the number of devices and flows remained relatively stable during the baseline period but exhibited a minor decline as the hurricane originated and approached. Notably, from August 25 through 27, 2021, as Hurricane Ida intensified into a Category 4 storm, there was a significant increase in the number of devices (+14.67%) and number of flows (+15.98%), suggesting people were mobilizing in response to the impending threat. After the peak, movements declined dramatically in terms of number of devices (-79.26%) and number of flows (-85.72%), indicating that population had completed preparation and was reducing movement to ensure safety. Following Ida's landfall on August 29 and its subsequent weakening, the pattern of movements showed a gradual recovery, returning to a semblance of normality by the middle of September.

Then daily global network properties, such as number of nodes, number of edges, average degree, density, clustering, diameter, and modularity, were calculated to assess the hurricane's influence on global network properties in the visitation network of places. These properties, as shown in the time series trends in Fig.4b, can be divided into two groups. The first group includes diameter and modularity, both exhibiting a rise and then fall during the hurricane period. The network diameter, a metric of the shortest path between the farthest nodes, expanded by +85.42% during the Ida and reverted to its pre-disaster level by September 2, signaling early recovery signs. Modularity quantifies the extent to which a network's nodes form distinct, densely connected subgroups. Increased modularity helps protect against widespread disturbances like natural hazards (Aquilué, Filotas et al. 2020). In our network of places, increased modularity (+32.75%) suggests more localized visitations within various clusters. These two properties highlight localized network connectivity. The initial surge in these properties before the Hurricane Ida's onset was likely due to intensified local travel for hurricane preparedness. However, during the hurricane, these properties significantly decreased, which was not due to a recovery of longer-distance visits, but rather because of a widespread cessation of travel, leading to a reduced flow within the overall network. The second group of properties involves number of nodes, number of edges, average degree, density, and clustering, all of which initially dropped before recovering. The average degree signals the mean number of connections each node has. Network density measures the tightness of these connections, and clustering refers to the extent to which nodes tend to cluster together. Two possible explanations exist for the trend exhibited by these properties. Initially, in the face of impending hurricane, people might break away from their regularly visited locations, moving towards resource-rich areas. This migration resulted in the observed drop in regular visit behavior. Once the hurricane receded, the properties' typical levels gradually recovered. The second explanation could be that the hurricane's approach directly diminished visitation and thus network connectivity, which in turn impacted the network's properties. These properties rebounded as the hurricane passed and connectivity was reestablished.

Overall, the global network properties reflect a population's different states of visitations in response to the hurricane. However, different properties imply different meanings, and it is not efficacious to examine the change in people's lifestyles in a unified perspective. Moreover, the global network properties reflect the overall location-based state of visitation from a higher-order perspective, and it is difficult to peek into the individual perspective to see how different lifestyles are coping with the natural hazards.

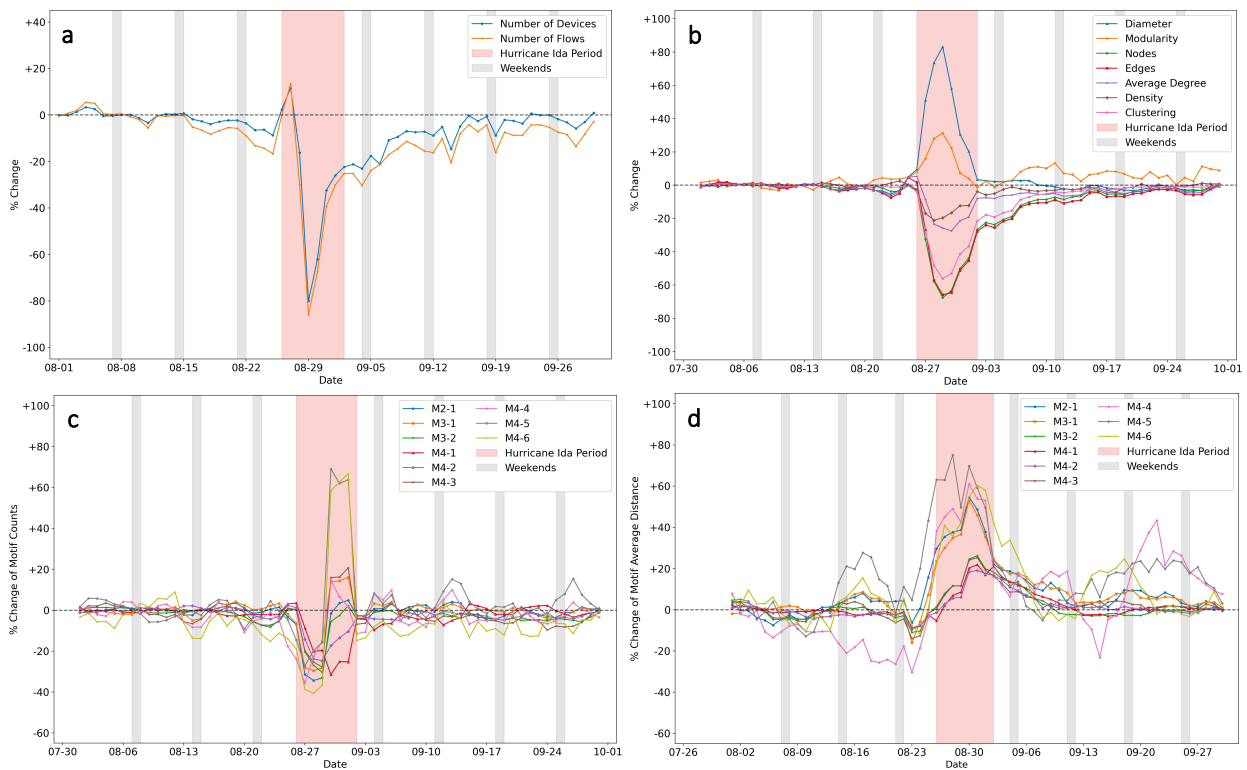

**Fig.4.** Temporal distribution of the network of places properties and motif metrics throughout the time duration before, during, and after Hurricane Ida. **a.** Temporal distribution of human movements. We extracted the number of flows and the number of devices as the basic statistics of raw human mobility data. **b.** Temporal distribution of global network properties in network of places. **c.** Temporal distribution of percentage change of motif frequency in the nine basic motifs. We used the count of motifs to calculate motif frequency. **d.** Temporal distribution of percentage change of motif proximity in the nine basic motifs. We used the average distance of motifs to calculate motif proximity.

### 4.2 Lifestyle fluctuations in the motifs

In this analysis, we decoded motifs to derive nine basic lifestyle patterns during the 61-day period from the networks of places. As shown in Table 1, M4-2, M3-2, and M4-1 had the highest frequencies, suggesting that they represent the most common lifestyle patterns. This commonality is likely attributed to these lifestyles facilitating easier access to locations and providing a greater variety of visitation routes. On the other hand, motifs like M4-4, M4-5, and M4-6 were observed less often, which could be due to less efficient structures within these lifestyle patterns, such as the hub-and-spoke layout observed in M4-6. Interestingly, the average distance across the various motifs was relatively consistent, suggesting that typical lifestyle movements involve locations within a 3- to 6-mile radius.

We then used motif frequency and motif proximity metrics to quantify the impact of the Hurricane Ida on the nine basic lifestyle patterns over time. We calculated the number of motifs as motif frequency and average distance as motif proximity in terms of percentage change from a baseline level detailed in the methodology section. We plotted the time series variation in Fig.4c and 4d. Fig.4c illustrates that the trend of motif frequency involves a sharp reduction followed by a sharp increase during the Hurricane Ida period. The sharp reduction took place primarily on August 26, with the greatest reduction in motif frequency falling between -20% to -40% across the nine basic motifs. The disruption in people's lifestyles, especially in terms of losing access to different locations due to the hurricane, was sudden rather than gradual. Despite advance disaster warnings, there seemed to be a preference for passive lifestyle adjustments with the sudden matter rather than incremental adjustments. Following the hurricane's landfall and its subsequent northward movement on August 29, a significant transition point occurred. After this date, a rebound of motif frequency was observed, particularly in M4-5 and M4-6. For example, M4-6 exhibited a significant rebound from -39.56% on August 28 to +67.23% on September 1. On the other hand, M4-1 and M4-2 showed a tendency towards modest and gradual recovery. By September, as the hurricane receded, most motifs began to show a reduction in their percentage change of frequency, except for M4-1 and M4-2, which continued a gradual recovery. Notably, by September 2, a stable trend in various motifs was observed, suggesting their adaptation ability to return to a normal state within a relatively short timeframe (8 days).

The varied disruption and rebound observed in frequency across different motifs highlight their differentiated representation of lifestyle resilience. The structures of M4-5 and M4-6 (linear chain and hub-and-spoke layout, respectively) are relatively fixed. To get access to any of the nodes, all the nodes must be recovered. Also, these motifs are probably to be involved in essential locations, such as hospitals, grocery stores, etc., and they should be recovered at the earliest possible time. On the other hand, M4-1 and M4-2 display more flexibility with a high level of interconnectedness between locations, allowing for various access routes and also possibly including some non-essential locations, making their recovery less urgent than others. Our analysis underscores the importance of examining specific locations for understanding differentiated changes in lifestyle patterns, which will be further explored in the upcoming section on assigning attributes to nodes.

We then examined how the Hurricane Ida influenced the nine basic lifestyle patterns in terms of spatial dispersion from the perspective of motif proximity, as shown in Fig.4d. First, starting from August 23, the proximity (average distance) of the motifs, such as M3-2, M4-1, M4-2, and M4-3, started to increase gradually, whereas the proximity of M2-1, M3-1, M4-4, M4-5, and M4-6 exhibited a more pronounced increase. The reason for the overall increasing trend is that people travelled to more distant locations that were not usually involved in their common routines in order to prepare ahead of time as part of their protective actions. As for differentiated increase trends, M2-1, M3-1, M4-4, M4-5, and M4-6 have a relatively fixed location-accessible structure, and thus show changes in overall distance due to the inclusion of these distant locations. Whereas M3-2, M4-1, M4-2, and M4-3 are relatively more flexible, benefitted from multiple access routes in maintaining shorter distances. In the post-hurricane period, a decrease in the proximity was observed, also falling into the two previously mentioned groups, each with varying rates and date of decrease. In addition, we can also observe that lifestyle patterns underwent a longer recovery period in terms of motif proximity (mid-September) compared to motif frequency (2 September). This longer recovery period is likely influenced by several factors including motif structure, spatial distance of locations, alterations in business hours of facilities, and the time required to repair property damage. Moreover, our results reveal considerable heterogeneity of proximity among the motifs, with motifs M4-4, M4-5, and M4-6 failing to regain stability by the end of the study period. This underscores the heterogeneity of recovery patterns for different lifestyles.

**Table 1.** Statistical summary of the nine basic lifestyle motifs.

| Motif Type | Motif frequency (Number) | Motif proximity (Average distance, miles) | Number of categories of attributed motifs |
|---|---|---|---|
| M2-1 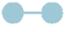 | 74,209 | 3.8938 | 249 |
| M3-1 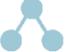 | 68,713 | 3.8789 | 1,143 |
| M3-2 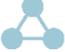 | 101,873 | 4.2476 | 1,701 |
| M4-1 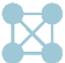 | 98,689 | 5.2332 | 5,570 |
| M4-2 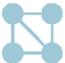 | 104,707 | 4.7668 | 5,668 |
| M4-3 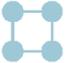 | 47,132 | 4.4819 | 4,208 |
| M4-4 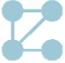 | 9,799 | 3.3149 | 1,064 |
| M4-5 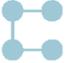 | 8,864 | 3.5032 | 1,379 |
| M4-6 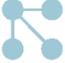 | 6,864 | 4.3516 | 1,317 |

While the nine basic motifs identified from the networks of places revealed the temporal and spatial characteristics of lifestyle signatures, we further differentiated these motifs based on node attributes to deeply explore the lifestyle heterogeneity. We assigned the 20 POI categories ( listed in Supplementary Table 1) to the nine basic motifs, generating a huge number of combinations of POI-based motifs, which we refer to as attributed motifs. Table 1 presents the number of categories for these attributed motifs. M4-2, M4-1, and M4-3 generated the largest number of attributed motifs, indicating a preference for these lifestyles due to their broad access to various locations. We also plotted the probability density distribution of frequency for attributed motifs in Supplementary Fig.2. We observed a significant heterogeneity in the frequency within different categories of attributed motifs, where a large number of frequencies is concentrated on a few attributed motifs (which means a few locations), while other attributes rarely appear in motifs.

Based on the above observation and for simplicity in further analysis, we focused on the top ten frequencies of attributed motifs within each of the nine basic motifs, as depicted in Fig.5. In each of the nine basic motifs, the frequencies of the top ten attributed motifs exceeded 10% of the totals. As for POI categories, essential ones were more frequent than non-essential ones. Essential POIs—grocery stores, healthcare facilities, and gasoline stations—are closely linked to everyday needs and are consistently found among almost all the top ten attributed motifs. Their presence in nearly every lifestyle pattern highlights the critical role in hurricane response. Conversely, despite being categorized as essential, educational and telecommunications services were not as prominent in the top ten attributed motifs. On the other hand, non-essential POIs, such as restaurants, clothing stores, financial investment services, automotive services, and amusement and recreation, showed up frequently in the top ten attributed motifs, indicating their significance in people's daily lifestyle patterns. These non-essential POIs, while not as vital as the above essential POIs for survival, are still integral to people's lifestyles and have possibly been undervalued in the past disaster management. Therefore, it is suggested that these non-essential locations should receive more consideration in disaster management strategies to reflect their role in everyday life.

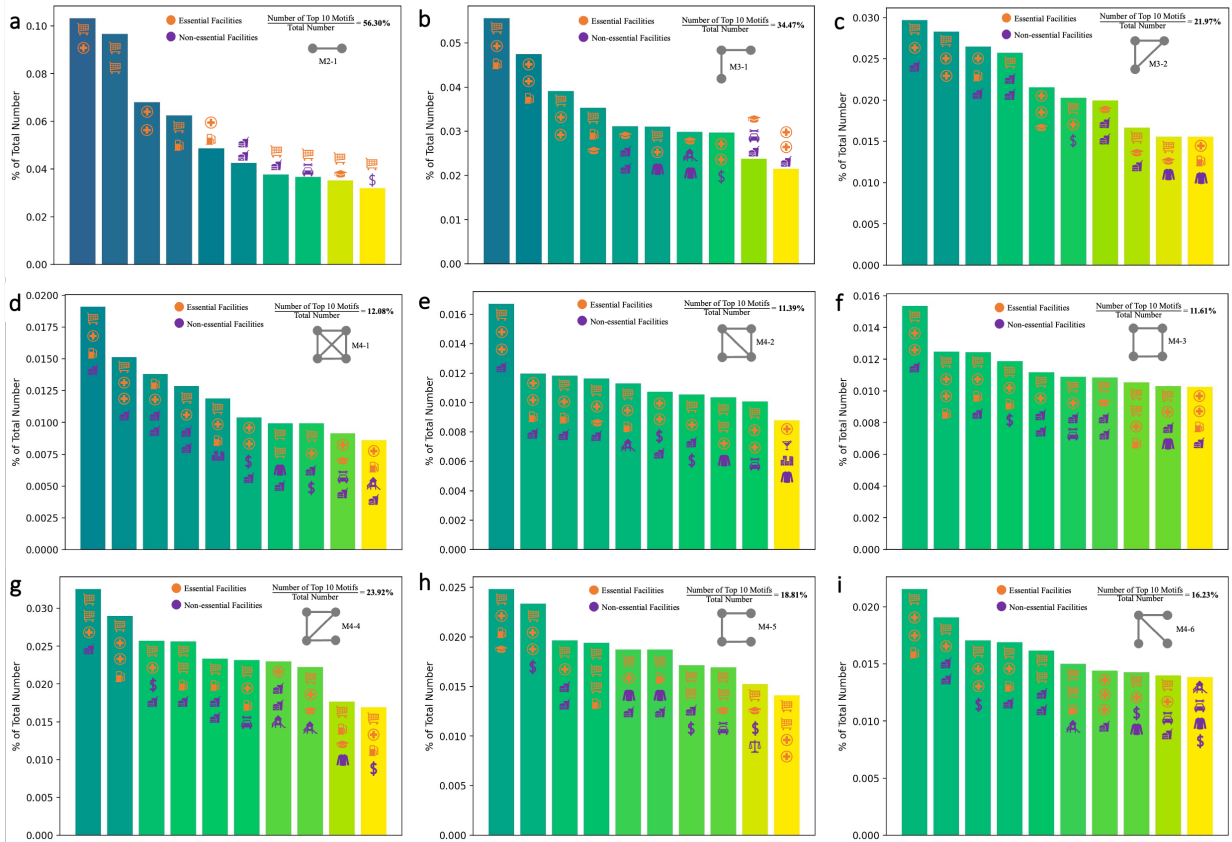

**Fig.5.** Top ten ranking of the attributed motifs in each of the nine basic motifs. Each bar is labeled with the icons of POI categories. The icons are arranged in a sequence from top to bottom in the bars, corresponding to the motif nodes' spatial configuration from left to right and top to bottom within the motif structure. The percentage is the proportion of the total number of motifs over the course of 61 days that are accounted within the top ten attributed motifs within that specific basic motif.

### 4.3 Lifestyle fluctuations in the attributed motif clusters

Upon analyzing the above top ten attributed motifs within each of the nine basic motifs, we identified recurring lifestyle patterns across these attributed motifs. For example, some attributed motifs characterized by frequent visits to offices (i.e., financial investment services, public administration, household, and real estate) display a commuting lifestyle pattern, while some focus on healthcare, such as drugstores or pharmacies, showing a healthcare-oriented lifestyle pattern. Therefore, we grouped and labeled the above top ten attributed motifs (90 attributed motifs in total) into four main lifestyle clusters: commute, healthcare, dining out, and youth. Fig.6 illustrates the connections among POI categories extracted from attributed motifs within the four lifestyle clusters. Supplementary Table 2 shows the exact composition of attributed motifs within each lifestyle cluster.

The commute lifestyle cluster accounted for 40.51% of the total motifs in terms of the number, indicating that a significant portion of lifestyles are related to commuting. This lifestyle cluster primarily involves essential POIs, such as grocery stores, gasoline stations, educational services, and non-essential POI,s like financial investment services, public administration, household and real estate, and automotive services. While POI categories such as healthcare and restaurants do exist within the commute lifestyle cluster, they are supplementary to the critical POIs that fundamentally define the lifestyle cluster (see Supplementary Table 2 for details). For example, visits to restaurants might occur in conjunction with visits to public administration or financial investment services, or a grocery store might be a stop after visits to educational services. The healthcare lifestyle cluster accounted for 34.01% of total motifs, and POIs related to healthcare (i.e., hospitals, pharmacies, and drugstores) were the focal locations from which visits radiate outward. The dining-out lifestyle cluster comprised 14.13% of total motifs, with restaurants being the central POI category. It reflects a lifestyle primarily oriented around eating out. Lastly, the young lifestyle cluster made up 11.34% of the total. This lifestyle is characterized by diverse consumption and recreational visitations, primarily involving amusement and recreation, clothing stores, and drinking places.

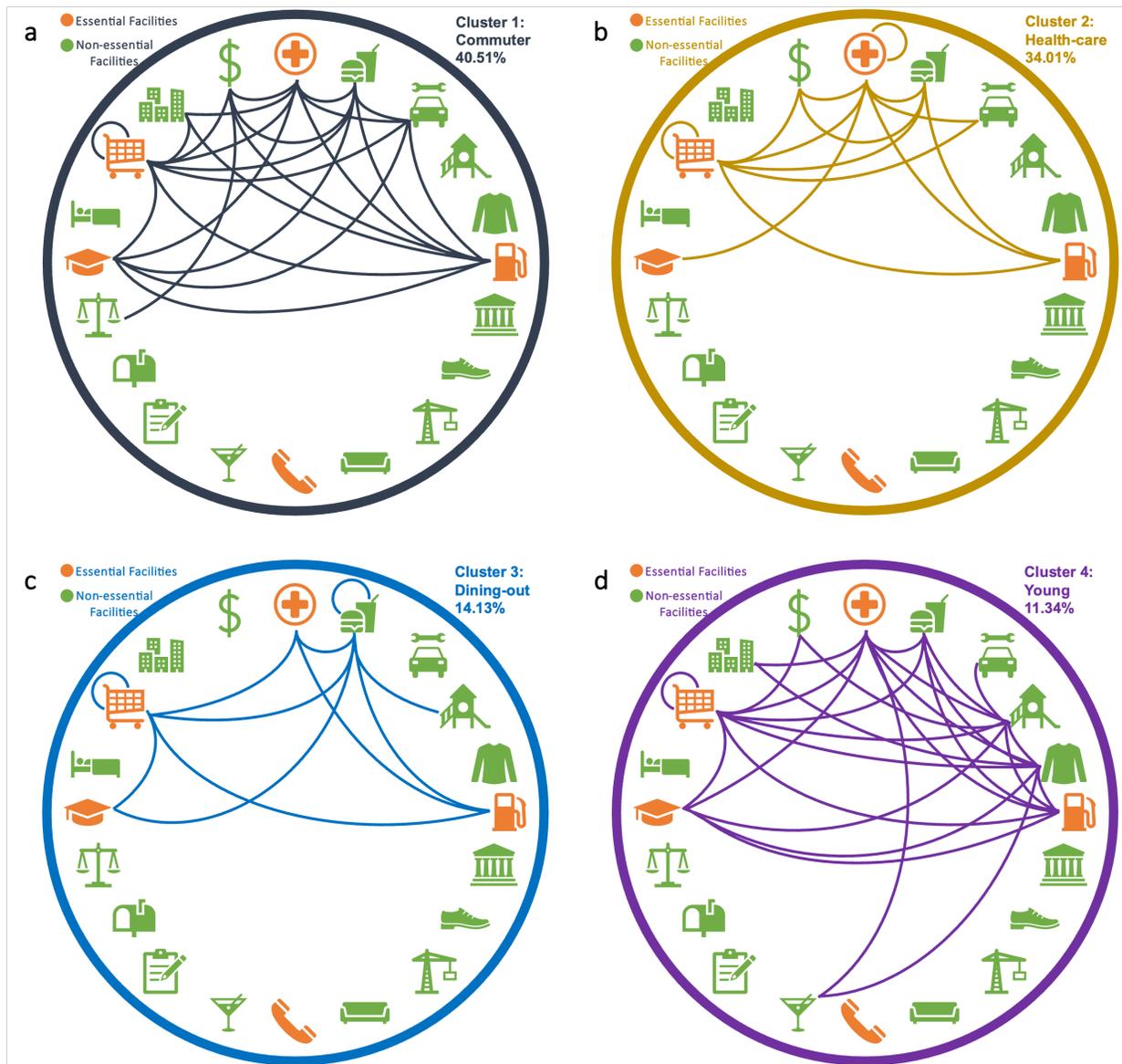

**Fig.6.** The four lifestyle clusters grouped by the top ten attributed motifs. Here, we deconstructed the attributed motifs belonging to certain lifestyle clusters into edges and nodes on either side of the edges, and then connected the corresponding nodes (icons in the cycles) with the edges in the lifestyle cluster cycles. Note that lifestyle clusters here do not show the real structure of the specific attributed motifs they contain. The exact composition of attributed motif in each lifestyle cluster can be viewed in Supplementary Table 2. Icons in the cycles represent POI categories. Names of each lifestyle cluster are noted on the top right corner. The percentage shown below the name is the proportion of the total number of motifs over 61 days that are accounted for by the number of the attributed motifs within that specific lifestyle cluster.

Then, we evaluated temporal dynamics of motif frequency and motif proximity for the four lifestyle clusters during Hurricane Ida. Fig.7 displays the temporal fluctuations of four lifestyle clusters throughout Hurricane Ida. Table 2 illustrates the varied maximum impact of Hurricane Ida on different lifestyle clusters and their respective recovery duration. In Fig.7a, the commute lifestyle cluster underwent a sequence of fluctuations in frequency: starting with a slight increase, followed by a sharp decrease, then a rapid rebound, and eventually leveling off. In contrast, the average commute distance showed an approximately opposite trend. Initially, there's a slight increase (+11.21%) in the frequency of commutes as people visited commuting-related POIs for preparing supplies and completing work before the hurricane's arrival. As the hurricane approached, there was a significant drop in frequency with the maximum decrease being -47.59% on Aug 29 due to the closure of related POIs. This, in turn, led to an increase in the average distance with the maximum increase reaching +26.98% on the same day as people visited further to access essential services. After the hurricane passed, there was a significant rebound in commuting visits, with the number not only recovering but also exceeding the pre-hurricane baseline by +22.72%. Concurrently, as the related POIs started to reopen, the necessity to visit longer distance diminishes, leading to a reduction in the average distance (-14.33%). Eventually, the motif frequency and motif proximity recovered to its steady state by Sep 5 (10 days).

The healthcare and dining-out lifestyle clusters exhibited similar patterns as the commute cluster in the metrics of motif frequency and motif proximity, though there were subtle but noteworthy differences among them. In the healthcare lifestyles, the increase slope and extent (+26.77%) in the frequency exceeded those observed in commute (+11.21%) in the initial stage. This is probably because the healthcare cluster encompasses essential POIs such as hospitals (e.g., hemodialysis centers and cancer diagnostic centers) that individuals cannot simply avoid. Thus, the frequency of visits may increase to make adequate preparations before the hurricane and may reduce only when the hurricane is landfalling. The percentage change in average distance visited to healthcare POIs (+40.05%) also exceed that for commute (+26.98%), indicating a necessity for population to travel to more distant hospitals in the event that the closest one is unavailable. Additionally, the relatively modest decrease of number (-32.05%) and a rapid temporal recovery to stability (September 3) in healthcare visits highlight the enduring importance of these facilities. Regardless of the stage of the hurricane, the population prioritizes returning to their usual healthcare routines as quickly as possible. In contrast, the dining-out lifestyle did not exhibit an initial increase in the motif frequency but experienced a longer recovery duration for motif frequency (11 days) and motif proximity (17 days), indicating its lesser significance relative to healthcare and commute in disaster preparedness. In addition, for the young-oriented lifestyle cluster, both frequency and proximity showed a significant decline (-19.98% and -28.65%, respectively) and prolonged recovery duration (12 days and 13 days, respectively) during the Hurricane Ida. This indicates that people tend to reduce their engagement in recreational activities as well as avoid traveling to far-off locations to ensure safety during the hurricane.

In addition, when examining the recovery of lifestyle clusters during the post-disaster period from September 2 through September 30, we observed that lifestyle clusters exhibited increased instability compared to the pre-disaster baseline period from August 1 through August 21. This is characterized by two elements: one is an increase in the randomness of both the number and average distance visited, and another is a diminished clarity in the weekly patterns typically seen during weekends and weekdays. This situation underscores the necessity for disaster management professionals to adopt a long-term perspective on the impacts of natural hazards on population lifestyles.

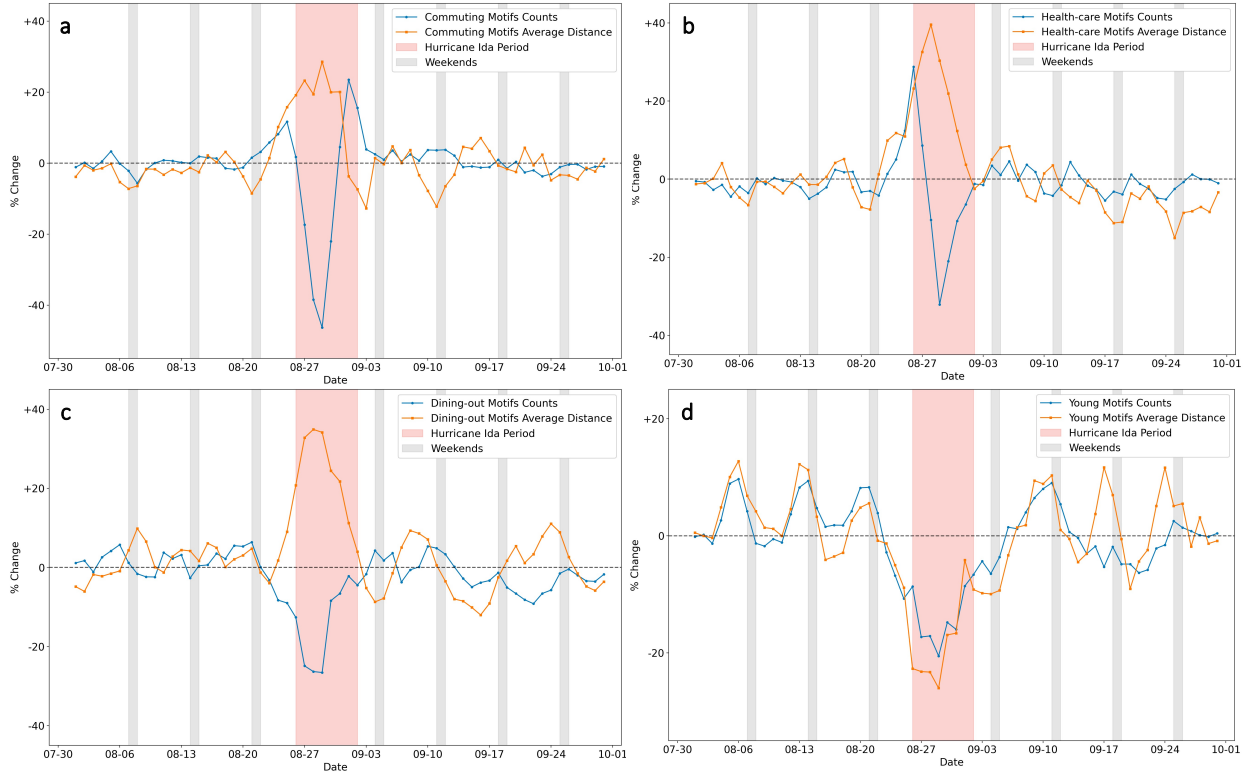

**Fig.7.** Temporal distribution of the four lifestyle clusters throughout Hurricane Ida. **a.** Temporal distribution of commute lifestyle cluster. **b.** Temporal distribution of healthcare lifestyle cluster. **c.** Temporal distribution of dining-out lifestyle cluster. **d.** Temporal distribution of young lifestyle cluster.

**Table 2.** The heterogenous impact of the Hurricane Ida on different lifestyle clusters and their recovery duration.

| Lifestyle cluster | Maximum impact (%) | | Recovery duration (days) | |
| --- | --- | --- | --- | --- |
| | Motif frequency | Motif proximity | Motif frequency | Motif proximity |
| Commute | -47.59 (Aug 29) | +26.98 (Aug 29) | 10 (Sep 5) | 10 (Sep 5) |
| Healthcare | -32.05 (Aug 29) | +40.05 (Aug 28) | 8 (Sep 3) | 8 (Sep 3) |
| Dining-out | -28.11 (Aug 29) | +33.37 (Aug 28) | 11 (Sep 6) | 17 Sep 12 |
| Young | -19.98 (Aug 29) | -28.65 (Aug 29) | 12 (Sep 7) | 13 (Sep 8) |

Note: We used the absolute maximum percentage change during the Ida period as the maximum impact for motif frequency and motif proximity. To determine the recovery durations, we looked for two consecutive days following the Ida period (September 2, 2021) where the percentage change remained within a ±5% threshold. A lifestyle cluster was considered recovered on the second of these two days, which served as the cutoff date. The period from the beginning of the Ida period (August 26, 2021) to this cutoff date was then calculated as the recovery duration. The specific dates when these values were observed are noted in parentheses.

## 5. Discussion and concluding remarks

The main idea of this study is to decode lifestyle signatures embedded in a population's visitation network of places and to evaluate fluctuations in lifestyle patterns to examine disaster impact and recovery in the affected populations. Departing from the existing literature that focuses primarily on evaluating physical infrastructure and the built environment to evaluate disaster impacts and recovery, this study emphasizes visitation to networks of places (POIs) and evaluates temporal fluctuations in sub-graph structures (motifs) to provide a more granular perspective into hazards impact on lifestyle signatures and their recovery. To this end, we utilized high-resolution human mobility data to construct the temporal human visitation networks in Louisiana. in the context of the 2021 Hurricane Ida. Using the constructed temporal network of places and location data, we investigated the temporal dynamics of network motifs, including their frequency and proximity during hazard perturbations. Our analysis uncovers hidden lifestyle patterns and fluctuations that contribute to the community resilience of population in the face of disaster-induced changes.

One of the major findings in this study is that we effectively delineated human lifestyle patterns through decoding network motifs. By analyzing the network of places, we distilled people's lifestyle patterns into nine basic motifs, each with distinct structures reflecting various human movement behaviors. Furthermore, we identified a broad spectrum of attributed motifs to reflect population's interactions with facilities utilizing the NAICS codes of the real-world locations. While these attributed motifs exhibited a diverse array and were marked by pronounced heterogeneity, we observed a frequency concentration on some few attributed motifs, indicating that certain POIs dominated, while others were infrequently visited. This pattern suggests a noticeable long-tail distribution in human lifestyles, aligning with findings from previous research (Ma, Li et al. , Di Clemente, Luengo-Oroz et al. 2018, Fan, Wu et al. 2024). In addition, we categorized the common features of these attributed motifs into four overarching lifestyle clusters: commute, healthcare, dining-out, and youth, which correspond with universal classifications of human behavior identified in earlier studies (Di Clemente, Luengo-Oroz et al. 2018, Fan, Wu et al. 2024). By matching individual-level mobility data and POI location data with the analysis of network motifs, our study offers a nuanced and granular portrayal of human lifestyles in cities during normal and crisis times. This approach advances beyond prior research, which typically segmented lifestyles into broad, simplistic categories like home and work (Lei, Chen et al. 2020), relied on coarse geographic scales such as census tracts (Delmelle and Nilsson 2021, Hsu, Liu et al. 2024), or used questionnaires for lifestyle assessment (Baker 2011, Josephson, Schrank et al. 2017).

By evaluating time series of motif frequency and motif proximity, this study captured the fluctuations in lifestyle patterns during disasters to better unravel community resilience dynamics. We constructed a temporal network model of daily human visitations to places and analyzed its properties based various metrics, including number of users, number of flows, global network properties, as well as the frequency and proximity of nine basic motifs and four lifestyle clusters. We employed a daily baseline calculation method to account for weekday and weekend lifestyle variations. This ensures precise measurement of percentage changes of our indicators throughout the study period, marking a significant enhancement in methodologies for lifestyles characterization research (Cox, Prager et al. 2011, Askitas, Tatsiramos et al. 2021). Accordingly, we successfully mapped the fluctuations in lifestyle patterns during Hurricane Ida, highlighting the population's response and recovery behaviors to the natural hazards' progression. In particular, the initial increase in human visitations, as indicated by the increase in frequency and proximity before Ida's landfall, was likely tied to preparedness actions, while the subsequent decline reflected the hunkering down during the peak of the hurricane. Subsequently, a gradual restoration was observed, signifying a return to pre-disaster lifestyles. The findings unveil the significance of examining lifestyle pattern fluctuations as an important, yet understudied aspect of community resilience dynamics in hazard events. We also identified that the resilience level varied across different lifestyle patterns. For example, M4-1 and M4-2 exhibited greater adaptability due to their high degree of location interconnectedness and diverse access routes, thereby minimizing disaster impacts on lifestyle patterns. Conversely, the healthcare lifestyle cluster showed heightened sensitivity to hazard-induced perturbations. These diverse resilience profiles across motifs and lifestyle clusters add depth to our understanding of how hazard events affect population life activities, moving us closer to better characterization and understanding of community resilience dynamics. In addition, our analysis indicates that global network properties are inadequate and fail to capture the true impacts of perturbations on human lifestyles. This finding implies that although numerous studies utilize global network properties to depict human behavior (Muchnik, Pei et al. 2013, Cimini and Sánchez 2014), careful consideration must be given to the choice of properties and their interpretation, and more attention should be paid to sub-structural dynamics in characterizing resilience properties in temporal networks.

Our study has effectively uncovered the varied patterns of how people prioritize visits to different locations during disasters. We observed that human lifestyles predominantly focused on a limited set of POIs, which can be classified into essential and non-essential categories. Essential POIs, such as grocery stores, health care facilities, and gasoline stations, were identified across nearly all the top ten attributed motifs, reflecting their fundamental role in daily necessities and disaster response. Moreover, some non-essential POIs, such as restaurants and automotive services, should not be overlooked, as they served as critical locations in the majority of lifestyles patterns. This suggests that effective disaster management should encompass a wide array of human needs to uphold societal well-being. This perspective challenges previous research that primarily focused on essential POIs (Esmalian, Coleman et al. 2022, Coleman, Liu et al. 2023, Jiang, Yuan et al. 2023), thus offering a broader understanding of lifestyle resilience. In detailing the four lifestyle clusters, our study pinpointed healthcare-related POIs, including hemodialysis centers and cancer diagnostic centers, as consistently critical. Despite the challenges posed by disasters, there was a concerted effort to ensure uninterrupted access to these important services. This aligns with prior studies that have underscored the significance of healthcare access during disasters (Liu and Mostafavi 2023, Yuan, Farahmand et al. 2023). Instead, for POIs such as entertainment facilities, we noted a high level of redundancy in people's lifestyles, indicating a more flexible attitude towards these services during disasters. These disparities in POI accessibility suggests the need for differentiated disaster management strategies, including tailored resource allocation and prioritization, to effectively sustain both essential services and the broader fabric of community life during emergencies.

Our study also reveals heterogeneity of different lifestyle patterns during the extended post-hurricane recovery period. We examined lifestyle restoration over a month following the Hurricane Ida. While some lifestyle clusters (e.g., healthcare) demonstrated rapid recovery to pre-hurricane conditions with strong rebound rates in motif frequency and proximity, the overall observations of lifestyle motifs and clusters indicated varied recovery durations for both motif frequency and proximity. This finding underscores heterogeneity in the recovery duration of lifestyle patterns and the variations in ways different subpopulations are impacted and recover in hazard events.

The findings obtained in this study have multiple scientific contributions and practical implications. First, this study enhances the understanding of human lifestyle resilience and adaptability during disasters by integrating high-resolution, location-specific human mobility data with the temporal dynamics of visitation network motifs. By dissecting human visitation networks into nine motifs and four lifestyle clusters, the findings contribute to interdisciplinary fields of urban science and risk management by providing valuable insights into the stability and regularity of urban lifestyles amidst crises. These insights can inform disaster managers and public officials about facilities that are critical for restoring community life activities after disasters. By understanding the constancy and patterns in how populations interact with urban spaces during emergencies, officials can better allocate resources, prioritize infrastructure development, and plan for disaster mitigation and recovery. Second, this study exhibits the heterogeneous nature of visitations to urban facilities across different disaster stages. The insight provides a deeper understanding of community resilience dynamics, which can thus inform disaster researchers about the relationship between different community characteristics and their lifestyle pattern fluctuations during hazard events. Furthermore, the metrics for quantifying disaster impacts on lifestyles and their recovery can provide emergency managers and public officials with new data-driven insight to monitor post-disaster impacts and recovery of populations. Last, our methods of lifestyle dynamics characterization offer a data-driven, quantitative methodological framework that can be applied to compare different cities and assess city-level metrics such as energy usage, equity, and access during crisis.

This work also has a number of limitations, which could be addressed in the future. One limitation is the inability to factor in social-demographic attributes due to privacy protection in the human mobility data, which could provide a more detailed understanding of social attributes that govern variations in lifestyles. Future studies could aim to integrate anonymized location-based data with additional datasets that could examine the influence of socio-demographic characteristics on human lifestyle patterns during disasters. Another limitation is that the study focuses primarily on lifestyle pattern characterization in the context of hazard events using one case study. The case study approach, while detailed, concentrates on single region during a single type of disaster, and a broader scope encompassing various population sizes, urban typologies, and geographic regions would be beneficial to fully understand human lifestyle disparities in different urban settings and during various disaster scenarios.

**Declarations of interest**

The authors declare no competing interests.

**Acknowledgement**

This material is based in part upon work supported by the National Science Foundation under Grant CMMI-1846069 (CAREER), the Texas A&M University X-Grant 699, and the Microsoft Azure AI for Public Health Grant. The authors also would like to acknowledge the data support from Spectus, Inc and SafeGraph, Inc.. Any opinions, findings, conclusions, or recommendations expressed in this material are those of the authors and do not necessarily reflect the views of the National Science Foundation, Texas A&M University, Microsoft Azure, Spectus, Inc., and SafeGraph, Inc..

**Data availability**

All data were collected through a CCPA- and GDPR-compliant framework and utilized for research purposes. The data that support the findings of this study are available from Spectus, Inc. and SafeGraph, Inc., but restrictions apply to the availability of these data, which were used under license for the current study. The data can be accessed upon request submitted on spectus.ai and safegraph.com. Other data we used in this study are all publicly available.

**Code availability**

The code that supports the findings of this study is available from the corresponding author upon request.

**Author contributions**

J.M.: Conceptualization, Methodology, Data curation, Formal analysis, Writing - Original draft. A.M.: Conceptualization, Methodology, Writing - Reviewing and Editing, Supervision, Funding acquisition.

**Additional information**

Supplementary information associated with this article can be found in the Supplementary Information document.

of pre-disaster owner, business and location characteristics." International Journal of Disaster Risk Reduction **23**: 25-35.

Kontokosta, C. E. and A. Malik (2018). "The Resilience to Emergencies and Disasters Index: Applying big data to benchmark and validate neighborhood resilience capacity." Sustainable cities and society **36**: 272-285.

Lei, D., X. Chen, L. Cheng, L. Zhang, S. V. Ukkusuri and F. Witlox (2020). "Inferring temporal motifs for travel pattern analysis using large scale smart card data." Transportation Research Part C: Emerging Technologies **120**: 102810.

Li, B. and A. Mostafavi (2022). "Location intelligence reveals the extent, timing, and spatial variation of hurricane preparedness." Scientific reports **12**(1): 16121.

Liu, C.-F. and A. Mostafavi (2023). "An equitable patient reallocation optimization and temporary facility placement model for maximizing critical care system resilience in disasters." Healthcare Analytics **4**: 100268.

Ma, J., B. Li and A. Mostafavi "Characterizing urban lifestyle signatures using motif properties in network of places." Environment and Planning B: Urban Analytics and City Science **0**(0): 23998083231206171.

Maeda, T. N., N. Shiode, C. Zhong, J. Mori and T. Sakimoto (2019). "Detecting and understanding urban changes through decomposing the numbers of visitors' arrivals using human mobility data." Journal of Big Data **6**: 1-25.

Milo, R., S. Shen-Orr, S. Itzkovitz, N. Kashtan, D. Chklovskii and U. Alon (2002). "Network motifs: simple building blocks of complex networks." Science **298**(5594): 824-827.

Muchnik, L., S. Pei, L. C. Parra, S. D. Reis, J. S. Andrade Jr, S. Havlin and H. A. Makse (2013). "Origins of power-law degree distribution in the heterogeneity of human activity in social networks." Scientific reports **3**(1): 1783.

Noszczyk, T., J. Gorzelany, A. Kukulska-Kozieł and J. Hernik (2022). "The impact of the COVID-19 pandemic on the importance of urban green spaces to the public." Land Use Policy **113**: 105925.

Podesta, C., N. Coleman, A. Esmalian, F. Yuan and A. Mostafavi (2021). "Quantifying community resilience based on fluctuations in visits to points-of-interest derived from digital trace data." Journal of the Royal Society Interface **18**(177): 20210158.

Prill, R. J., P. A. Iglesias and A. Levchenko (2005). "Dynamic properties of network motifs contribute to biological network organization." PLoS biology **3**(11): e343.

Rajput, A. A. and A. Mostafavi (2023). "Latent sub-structural resilience mechanisms in temporal human mobility networks during urban flooding." Scientific Reports **13**(1): 10953.

Ronco, M., J. M. Tárraga, J. Muñoz, M. Piles, E. S. Marco, Q. Wang, M. T. M. Espinosa, S. Ponserre and G. Camps-Valls (2023). "Exploring interactions between socioeconomic context and natural hazards on human population displacement." Nature Communications **14**(1): 8004.

SafeGraph (2023). https://www.safegraph.com/.

Spectus (2023). https://spectus.ai/.

Su, R., E. C. McBride and K. G. Goulias (2020). "Pattern recognition of daily activity patterns using human mobility motifs and sequence analysis." Transportation Research Part C: Emerging Technologies **120**: 102796.

Thombre, A. and A. Agarwal (2021). "A paradigm shift in urban mobility: Policy insights from travel before and after COVID-19 to seize the opportunity." Transport Policy **110**: 335-353.

Toole, J. L., C. Herrera-Yaqüe, C. M. Schneider and M. C. González (2015). "Coupling human mobility and social ties." Journal of The Royal Society Interface **12**(105): 20141128.

Wang, B., B. P. Loo, F. Zhen and G. Xi (2020). "Urban resilience from the lens of social media data: Responses to urban flooding in Nanjing, China." Cities **106**: 102884.

Yang, Y., A. Heppenstall, A. Turner and A. Comber (2019). "A spatiotemporal and graph-based analysis of dockless bike sharing patterns to understand urban flows over the last mile." Computers, Environment and Urban Systems **77**: 101361.

Yin, J. and G. Chi (2021). "Characterizing people's daily activity patterns in the urban environment: a mobility network approach with geographic context-aware Twitter data." Annals of the American Association of Geographers **111**(7): 1967-1987.

Yuan, F., H. Farahmand, R. Blessing, S. Brody and A. Mostafavi (2023). "Unveiling dialysis centers' vulnerability and access inequality during urban flooding." Transportation Research Part D: Transport and Environment **125**: 103920.

Yuan, F. and R. Liu (2020). "Mining social media data for rapid damage assessment during Hurricane Matthew: Feasibility study." Journal of Computing in Civil Engineering **34**(3): 05020001.
23

Supplementary Information for

*Decoding the Pulse of Community during Disasters: Resilience Analysis Based on Fluctuations in Latent Lifestyle Signatures within Human Visitation Networks*

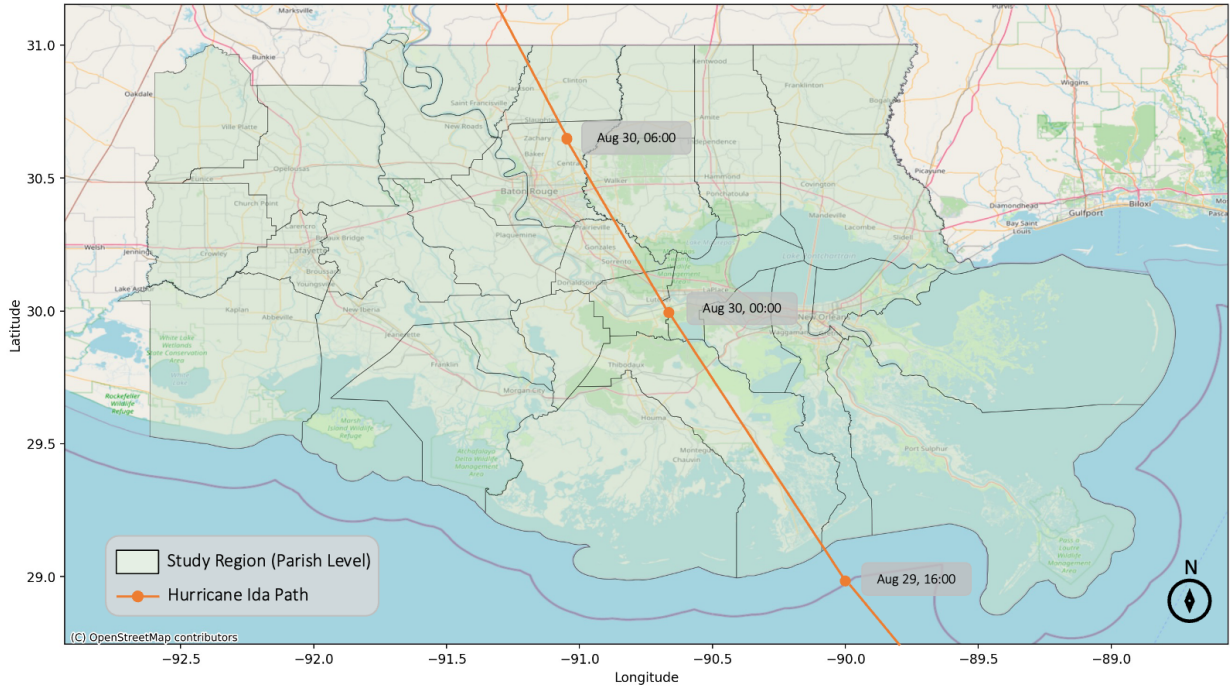

**Fig 1.** The moving path of Hurricane Ida and the 30 parishes in this study

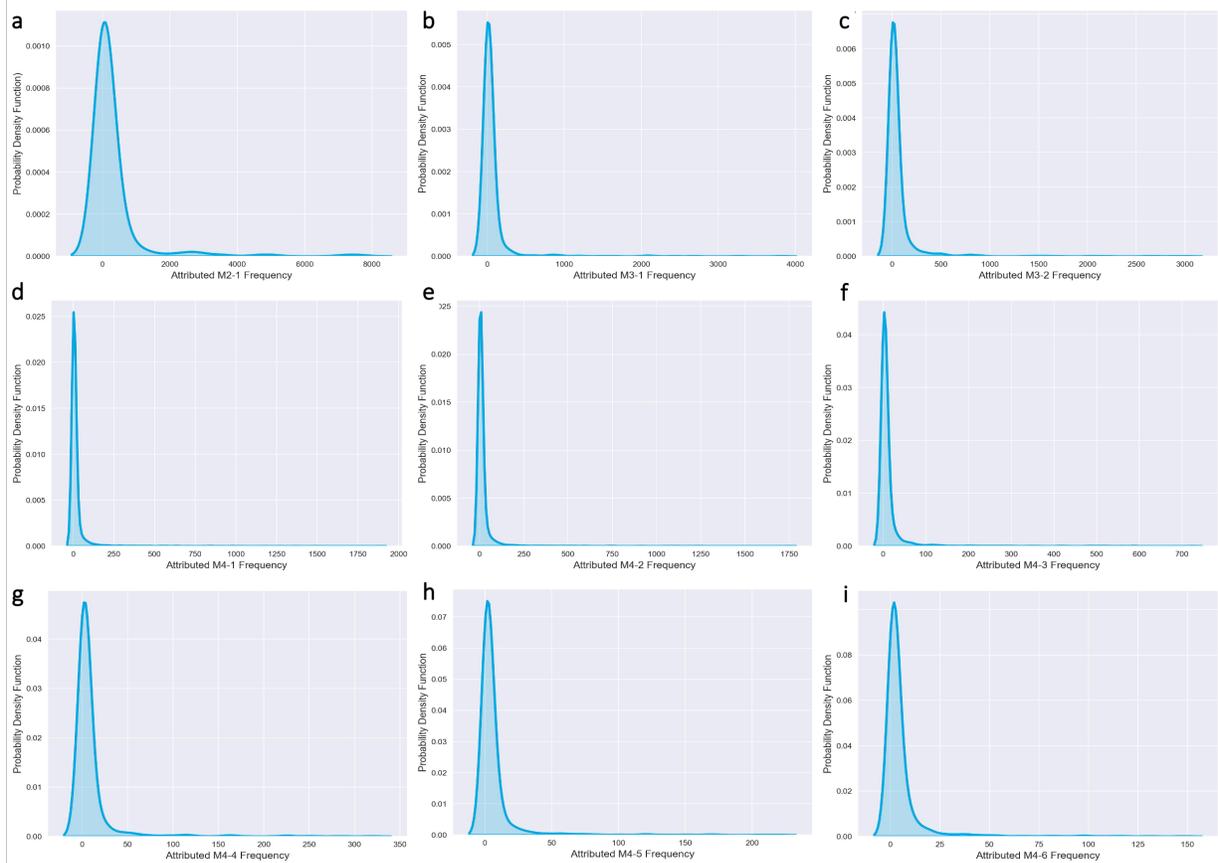

**Fig 2.** Probability density distribution of frequency for attributed motifs in each of the 9 basic motifs

| Category | Facility Name | NAICS Last 4 Digit Code |
|---|---|---|
| **Table 1.** The summary of essential/non-essential POI facilities and corresponding NAICS codes | | |
| Essential Facilities | Health Care | 4461, 6211, 6212, 6213, 6214, 6215, 6216, 6219, 6221, 6222, 6223, 6231, 6233, 6244, 8121, 8122, 8129 |
| | Grocery Stores | 3118, 3399, 4239, 4242, 4246, 4249, 4431, 4451, 4452, 4483, 4512, 4522, 4523, 4531, 4532, 4533, 4539, 4543, 8114 |
| | Gasoline Stations | 4471 |
| | Telecommunications Carriers | 5173 |
| | Educational Service | 5182, 5416, 5419, 6111, 6112, 6113, 6114, 6115, 6116 |
| Non-essential Facilities | Restaurants | 3119, 7223, 7225 |
| | Financial Investment Service | 5221, 5222, 5223, 5239, 5614 |
| | Household and Real Estate | 3352, 4009, 4236, 4237, 4442, 4931, 5311, 5312, 5313, 5322, 5323, 5616, 8123 |
| | Clothing Stores | 3159, 4481 |
| | Building Construction Service | 2361, 2381, 2382, 2383, 3271, 3272, 3323, 3325, 3328, 4233, 4441, 5324, 5617 |
| | Amusement and Recreation | 4511, 7111, 7112, 7113, 7131, 7132, 7139, 5121, 5122 |
| | Automotive Service | 3361, 4231, 4234, 4238, 4411, 4412, 4413, 4842, 4853, 5321, 7212, 8111, 8112 |
| | Public Administration | 2211, 3231, 4821, 4832, 4851, 4852, 4859, 4884, 5152, 5191, 5411, 5412, 5418, 5511, 5613, 5621, 5622, 5629, 6242, 8133, 9221, 9261 |
| | Insurance Service | 5241, 5242 |
| | Shoe Stores | 4482 |
| | Drinking Places | 3121, 4453, 7224 |
| | Furniture Stores | 4421, 4422 |
| | Museums and Historical Sites | 7121 |
| | Traveler Accommodation | 5615, 7211 |
| | Postal Services | 4841, 4881, 4885, 4911, 4921 |

**Table 2.** The composition of attributed motifs within the four lifestyle clusters.

| Lifestyle Type | M2-1 | M3-1 | M3-2 | M4-1 | M4-2 | M4-3 | M4-4 | M4-5 | M4-6 |
|---|---|---|---|---|---|---|---|---|---|
| Commute | | | | | | | | | |
| Healthcare | | | | | | | | | |
| Dining-out | | | | | / | | | | |

| | | | | | | | | | | |
|---|---|---|---|---|---|---|---|---|---|---|
| | | 🛒—🍔 | / | 🎓—🛍️/🍔 | 🛒—⊕/🍔—🍔 | / | 🛒—🎓/🍔—🍔 | ⊕—🛍️/🍔—🛝 | / | 🛒—🛒/🍔—🍔 |
| Young | | / | 🛒—⊕/👕 | 🛒—🎓/👕 | 🛒—🛒/👕—🍔 | 🛒—⊕/⛽—🛝 | 🛒—⊕/👕 | 🛒—⊕/🎓—🛝 | 🛒—⊕/👕—🍔 | 🛒—🛒/⛽—🏠 |
| | | / | 🎓—🛝/👕 | ⊕—⛽/👕 | ⊕—⛽/🛝—🍔 | 🛒—🛒/⊕—👕 | / | 🛒—⛽/🎓—🏠 | 🛒—⛽/👕—🍔 | 🛒—⊕/$—👕 |
| | | / | / | / | / | ⊕—🍸/🏙️—👕 | / | / | / | 🛝—🚗/🏠—$ |

Note:
🟠 Essential Facilities
🟣 Non-essential Facilities